\documentclass[aps,showpacs,twocolumn]{revtex4}

\usepackage{graphics}
\usepackage{epsfig}

\def\ov#1{\overline{#1}}

\def\wh#1{\widehat{#1}}

\def\hs{\heartsuit}

\newcommand{\bc}{\begin{center}}
\newcommand{\ec}{\end{center}}
\newcommand{\bt}{\begin{tabbing}}
\newcommand{\et}{\end{tabbing}} 
\newcommand{\be}{\begin{eqnarray*}}
\newcommand{\ee}{\end{eqnarray*}}
\newcommand{\bs}{\begin{slide}}
\newcommand{\es}{\end{slide}}

\begin{document}

\title{Analysis of ray phase-space recirculation in an extended Budden problem}

\author{A.~J.~Brizard}
\affiliation{Department of Chemistry and Physics, Saint Michael's College, Colchester, VT 05439} 

\author{A.~N.~Kaufman}
\affiliation{Lawrence Berkeley National Laboratory, University of California, Berkeley, CA 94720}

\author{E.~R.~Tracy}
\affiliation{Department of Physics, College of William and Mary, Williamsburg, VA 23187-8795}

\begin{abstract}
A three-wave Budden model with two resonance layers is constructed that allows recirculation of energy fluxes along a quadrangle in ray phase space. The transmission, reflection, and conversion coefficients for this extended Budden problem are calculated by ray phase-space methods and the modular-eikonal approach. Analytical and numerical results show that all coefficients exhibit interference effects that depend on an interference phase calculated from the coupling constants and the area enclosed by the quadrangle.
\end{abstract}

\begin{flushright}
March 29, 2007
\end{flushright}


\maketitle

\begin{figure}
\epsfysize=3in
\epsfbox{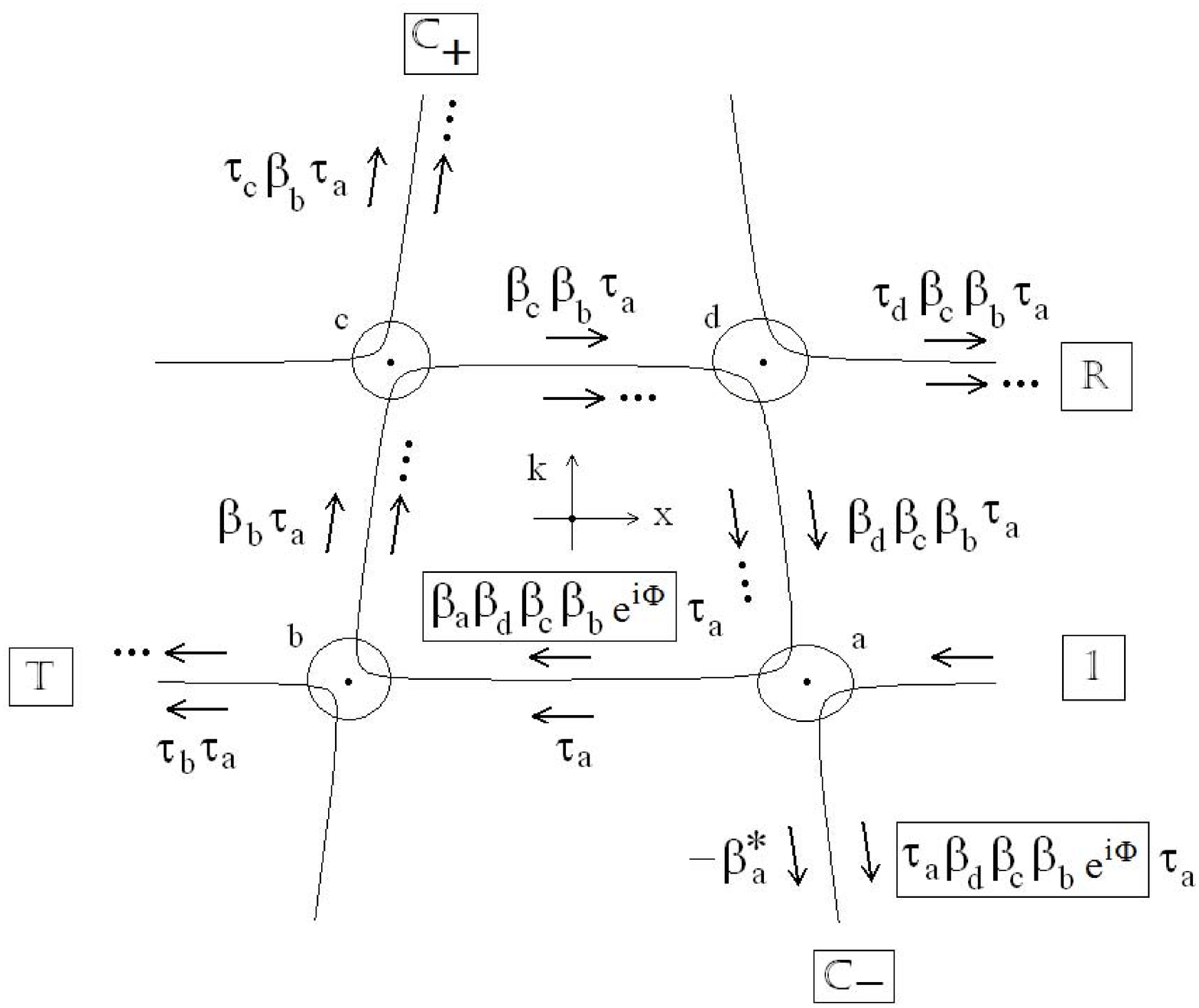}
\caption{Counter-$k$-propagating rays $B_{\pm}$ interacting with $x$-propagating ray $A$. Solid lines are dispersion curves representing Eq.~(\ref{eq:detD}). The four conversion regions $(a,b,c,d)$ form a quadrangle in ray phase space and arrows show the direction of wave-energy flow. The net propagation phase $\Phi = \oint k\,dx$ is shown only in the boxed expressions. For each unit of energy carried by the incoming left-propagating ray $A$ (from $x = +\,\infty$) introduced at conversion region $a$, the outgoing left-propagating ray $A$ carries the transmitted energy flux ${\sf T}$ (left of conversion region $b$), the outgoing up-propagating ray $B_{+}$ carries the converted energy flux ${\sf C}_{+}$ (above conversion region $c$), the outgoing right-propagating ray $A$ carries the reflected energy flux ${\sf R}$ (right of conversion region $d$), and the outgoing down-propagating ray $B_{-}$ carries the converted energy flux ${\sf C}_{-}$ (below conversion region $a$). Conservation of energy flux requires that ${\sf T} + {\sf C}_{+} + {\sf C}_{-} + {\sf R} = 1$.}
\label{fig:neg_B}
\end{figure}

The process of mode conversion, involving the interaction between two distinct waves propagating in a magnetized plasma, represents an important mechanism by which heating and current drive by radio frequency may lead to a burning plasma. In a standard single-crossing mode-conversion scenario (see conversion region $a$ in Fig.~\ref{fig:neg_B}), an incident ray from a primary ($x$-propagating) wave propagates on its dispersion surface until it crosses transversely the dispersion surface of a secondary ($k$-propagating) wave. Mode conversion occurs when some of the energy flux associated with the primary wave is converted to energy flux for the secondary wave, which allows a secondary-wave ray to propagate away from the conversion region. The remaining energy flux associated with the primary wave is transmitted across the conversion region and allows a primary-wave ray to propagate away from the conversion region. The conservation of energy (flux) dictates that the sum of the outgoing energy fluxes is equal to the incoming energy flux.

The classic Budden problem is a double-conversion process in ray phase space \cite{TK_93} (see conversion regions $b$ and 
$c$ in Fig.~\ref{fig:neg_B}), whereby a primary incoming wave is converted to a secondary wave at a first conversion region ($b$), which then propagates to a second conversion region ($c$) where it is converted to an outgoing (reflected) primary wave. Using ray phase-space methods and the modular-eikonal approach \cite{Liang}, we extend the standard Budden problem by adding a second double-conversion process (see conversion regions $a$ and $d$ in Fig.~\ref{fig:neg_B}), and a second resonance layer, associated with a third wave possibly supported by an energetic-particle population in an inhomogeneous magnetized plasma. The extended three-wave Budden problem introduces the possibility of a leaky cavity being created in ray phase space that allows recirculation of energy fluxes.

The modular-eikonal analysis of energy-flux recirculation around a leaky phase-space cavity in ray phase space was first performed by Brizard {\it et al.} \cite{Brizard_1,Brizard_2}, who considered the linear interaction between two waves with dispersion curves modeled by intersecting parabolas. In this earlier work, the modular-eikonal approach was successfully applied to the double-crossing mode conversion process whereby a ray of one wave punctured the dispersion surface of another wave twice due to ray curvature (corresponding to the case of a near-tangential crossing). The ray phase-space analysis revealed the existence of four mode-conversion scenarios based on whether the wave energies had equal or opposite signs, and whether the wave rays were co-propagating or counter-propagating in ray phase space (i.e., whether the signs of $dk/dt$ were equal or opposite). While all four scenarios yielded transmission and conversion coefficients exhibiting interference effects associated with the area enclosed by the intersecting parabolas and the strength of the coupling constants at the two conversion points, only two scenarios involved recirculation of energy fluxes.

The present work considers a modification of the three-wave Budden problem studied by Liang {\it et al.} \cite{Liang}, who considered the mode conversion in two-dimensional ray phase space $(x,k)$ between an $x$-propagating wave with two positive-energy waves that are co-propagating along the $k$-axis. The extended three-wave Budden problems, considered here and in Ref.~\cite{Liang}, introduce four conversion regions $(a,b,c,d)$ (instead of two for the classic Budden problem) that form the vertices of a quadrangle in ray phase space produced by the pair-wise intersections of 4 uncoupled rays associated with three waves (see Fig.~\ref{fig:neg_B}). The modular-eikonal analysis of Liang {\it et al.} \cite{Liang} showed that the reflection coefficient and one of the conversion coefficients exhibited interference effects that depended on the area enclosed by the quadrangle and the strength of the coupling constants at the four vertices. In addition, Liang {\it et al.} \cite{Liang} obtained excellent agreement between direct numerical integration of the coupled-wave equations and analytical results, thus confirming the validity of the modular-eikonal approach.

The purpose of the present paper is to use the modular-eikonal approach to study the extended three-wave Budden problem in which two counter-propagating positive-energy waves are each interacting linearly with an $x$-propagating wave. This scenario allows us to investigate the validity of the modular-eikonal approach when energy fluxes are allowed to recirculate around a quadrangle in ray phase space.

We begin with the extended three-wave Budden model
\begin{equation}
\left( \begin{array}{ccc}
D_{A}(x,\wh{k})        & \eta_{+}   & \eta_{-} \\
\eta_{+}^{*}        & D_{+}(x,\wh{k})   & 0 \\
\eta_{-}^{*}        & 0       & D_{-}(x,\wh{k}) 
\end{array} \right) \left( \begin{array}{c}
A(x) \\
B_{+}(x) \\
B_{-}(x)
\end{array} \right) \;=\; 0,
\label{eq:extended_B}
\end{equation}
where the $x$-propagating wave $A$ is represented by its dispersion function $D_{A}(k) = k_{0}^{2} - k^{2}$ (e.g., a magnetosonic wave) and the two $k$-propagating waves $B_{\pm}$ are represented by their dispersion functions $D_{\pm}(x,k) = (\sigma \pm x) - \gamma\,k$. The bulk-ion and energetic-particle positive-energy waves are represented by the 
$B_{+}$ and $B_{-}$ waves, respectively, and the standard Budden problem is recovered with $\eta_{-} = 0$. In 
Eq.~(\ref{eq:extended_B}), $\wh{k} = -\,i\,d/dx$ and the strengths of the mode conversions $A \leftrightarrow 
B_{\pm}$ are governed by the coupling constants $\eta_{\pm}$. The upward and downward $k$-propagating waves (separated by the distance $2\,\sigma$ on the $x$-axis) are allowed to propagate (in $x$-space) by setting the parameter 
$\gamma > 0$. By using the $3\times 3$ dispersion matrix ${\sf D}(x,k)$ of Eq.~(\ref{eq:extended_B}), the dispersion curves (shown in Fig.~\ref{fig:neg_B}) are obtained from the dispersion relation 
\begin{equation}
{\rm det}\,{\sf D} \;=\; D_{A}\,D_{+}\,D_{-} \;-\; |\eta_{+}|^{2}\,D_{-} \;-\; |\eta_{-}|^{2}\,D_{+} \;=\; 0,
\label{eq:detD}
\end{equation}
and the uncoupled dispersion curves cross in ray phase space at the conversion points $(x_{(a,d)}, k_{(a,d)}) = (\sigma \pm \gamma\,k_{0}, \; \mp k_{0})$ and $(x_{(c,b)}, k_{(c,b)}) = (-\,\sigma \pm \gamma\,k_{0},\; \pm k_{0})$. The energy-flux propagation is represented by the Hamilton equations $(\dot{x} = -\;\partial_{k}D$, $\dot{k} = \partial_{x}D$), which yield $(\dot{x}_{A},\; \dot{k}_{A}) = (\pm\,2,\; 0)$ and $(\dot{x}_{\pm},\; \dot{k}_{\pm}) = (\gamma,\;\pm\,1)$, indicated by arrows in Fig.~\ref{fig:neg_B}. We note that the symmetric-quadrangle dispersion curves associated with the extended Budden model (\ref{eq:extended_B}) facilitate the calculations done here, while the use of an asymmetric-quadrangle model (based on generic dispersion functions $D_{\pm}$ with $\dot{x}_{-} \neq \dot{x}_{+}$ and $|\dot{k}_{-}| \neq |\dot{k}_{+}|$) would not qualititatively change our results (as will be discussed below). 

The coupled differential equations of the extended Budden model (\ref{eq:extended_B}):
\begin{eqnarray}
\frac{d^{2}A(x)}{dx^{2}} \;+\; k_{0}^{2}\;A(x) & = & -\,\sum_{j = \pm}\;\eta_{j}\;B_{j}(x) \nonumber \\
 &  & \label{eq:ODE_Budden} \\
i\,\gamma\;\frac{dB_{\pm}(x)}{dx} \;+\; (\sigma \pm x)\;B_{\pm}(x) & = & -\,\eta_{\pm}^{*}\;A(x) \nonumber
\end{eqnarray}
satisfy the wave energy-flux conservation law 
\begin{equation}
\frac{d}{dx} \left( J_{A}(x) \;+\; \sum_{j = \pm}\;J_{j}(x) \right) \;=\; 0, 
\label{eq:action_con}
\end{equation}
where $J_{A}(x) \equiv {\rm Im}(A^{*}\,dA/dx)$ and $J_{\pm}(x) \equiv \frac{1}{2}\,\gamma |B_{\pm}|^{2}$. The choice of the $x$-space propagation parameter $\gamma$ for waves $B_{\pm}$ does not affect the results of the modular-eikonal approach \cite{Liang} presented for our symmetric-quadrangle model. When $\gamma = 0$, we replace $B_{\pm}$ with $-\,[\eta_{\pm}^{*}/(\sigma \pm x)]\,A$ in the wave equation for $A(x)$ and we obtain the extended Budden equation
\begin{equation}
\frac{d^{2}A}{dx^{2}} \;+\; \left[\; k_{0}^{2} \;-\; \left( \frac{|\eta_{+}|^{2}}{\sigma + x} \;+\; 
\frac{|\eta_{-}|^{2}}{\sigma - x} \right) \;\right] A \;=\; 0,
\label{eq:ext_Budden}
\end{equation}
which generalizes the standard Budden equation by considering two distinct resonance layers (at $x = \pm\,\sigma$) instead of a single resonance. A similar extension of the standard Budden problem associated with the presence of sheared flows in magnetized plasmas leads to three resonance layers and a phase-space cavity \cite{triplicate}.

We solve the extended Budden problem (\ref{eq:extended_B}) by using the modular-eikonal analysis \cite{Liang,Brizard_1}, which involves the following single-crossing amplitude transmission and conversion coefficients $(\tau_{a},\beta_{a}) = (\tau_{d},\beta_{d}) = (\tau_{-},\beta_{-})$ and $(\tau_{b},\beta_{b}) = (\tau_{c},\beta_{c}) = (\tau_{+},\beta_{+})$, where
\begin{equation}
\left. \begin{array}{rcl}
\tau_{\pm} & = & \exp\left(-\,\pi\;|\ov{\eta}_{\pm}|^{2}\right) \\
 &  & \\
\beta_{\pm} & = & \sqrt{1 - \tau_{\pm}^{2}}\; \exp (i\,\varphi_{\pm})
\end{array} \right\},
\label{eq:trans_conv}
\end{equation}
with the conversion phases $\varphi_{\pm} \equiv {\rm arg}[\Gamma(i|\ov{\eta}_{\pm}|^{2})]$ and $|\ov{\eta}_{\pm}|^{2} \equiv |\eta_{\pm}|^{2}/(2k_{0})$, where $2k_{0} = |\{D_{A},\, D_{\pm}\}|$ denotes the ray phase-space Poisson bracket of the dispersion functions evaluated at the conversion points. The three-wave conversion scenario shown in Fig.~\ref{fig:neg_B} exhibits a leaky cavity feature in ray phase space that allows energy flux to recirculate around the quadrangle $abcd$. To represent this process, we introduce the complex-valued recirculation coefficient (denoted by the heart symbol)
\begin{equation}
\hs \;\equiv\; \beta_{d}\,\beta_{c}\,\beta_{b}\,\beta_{a}\;e^{i\Phi} \;=\; |\beta_{+}|^{2}\,|\beta_{-}|^{2}\;e^{i\Psi},
\label{eq:recir_plus}
\end{equation}
with the interference phase
\begin{equation}
\left. \begin{array}{rcl}
\Psi & \equiv & \Phi \;+\; 2\,(\varphi_{+} \;+\; \varphi_{-}) \\
\Phi & \equiv & 4\,k_{0}\sigma \;-\; 2\,\left(|\ov{\eta}_{+}|^{2} + |\ov{\eta}_{-}|^{2}\right)\;\ln(4\,k_{0}\sigma)
\end{array} \right\}
\label{eq:Psi_counter}
\end{equation}
expressed as the sum of the propagation phase $\Phi \equiv \oint k\,dx$ (expressed as the sum of the area $4\,k_{0}
\sigma$ enclosed by the uncoupled quadrangle $abcd$ and corrections to the uncoupled eikonal phases due to finite coupling constants $\eta_{\pm} \neq 0$) and the net conversion phase ${\rm arg}(\beta_{a}\beta_{b}\beta_{c}\beta_{d}) = 2\,(\varphi_{+} + \varphi_{-})$. Note that the magnitude of the recirculation coefficient $|\hs|$ increases from 0 to 1 as the coupling constants $|\ov{\eta}_{\pm}|$ go from 0 to infinity. 

All coefficients $({\sf T}, {\sf C}_{\pm}, {\sf R})$ are calculated by adding the amplitudes of each successive completed circuit in Fig.~\ref{fig:neg_B} and, thus, they exhibit interference effects that depend on $\Psi$. These interference effects enter through the recirculation coefficient (\ref{eq:recir_plus}) in two different ways in Fig.~\ref{fig:neg_B}. First, successive powers of $\hs$ are added to the zeroth-order amplitudes for the transmission 
${\sf T}$, conversion ${\sf C}_{+}$, and reflection ${\sf R}$ coefficients; see the recirculation coefficient $\hs$ in the boxed expression in 
Fig.~\ref{fig:neg_B} between regions $a$ and $b$. Hence, the net transmission coefficient ${\sf T} \equiv |A_{out}^{b}|^{2}/|A_{in}^{a}|^{2}$ is
\begin{eqnarray}
{\sf T} & = & \left| \tau_{b}\,\tau_{a}\; \left( 1 \;+\; \hs \;+\; \hs^{2} \;+\; \cdots \right) \right|^{2} \nonumber \\
 & \equiv & \tau_{+}^{2}\tau_{-}^{2}\;|1 \;-\; \hs|^{-2},
\label{eq:T_counter_plus}
\end{eqnarray}
the net conversion coefficient ${\sf C}_{+} \equiv |B_{+ out}^{c}|^{2}/|A_{in}^{a}|^{2}$ to the upward $k$-propagating wave is
\begin{eqnarray}
{\sf C}_{+} & = & \left| \tau_{c}\,\beta_{b}\,\tau_{a}\; \left( 1 \;+\; \hs \;+\; \hs^{2} \;+\; \cdots \right) 
\right|^{2} \nonumber \\
 & = & \tau_{+}^{2}\,(1 - \tau_{+}^{2})\,\tau_{-}^{2}\;|1 \;-\; \hs|^{-2},
\label{eq:C1_counter_plus}
\end{eqnarray}
the net reflection coefficient ${\sf R} \equiv |A_{out}^{d}|^{2}/|A_{in}^{a}|^{2}$ is
\begin{eqnarray}
{\sf R} & = & \left| \tau_{d}\,\beta_{c}\,\beta_{b}\,\tau_{a}\; \left( 1 \;+\; \hs \;+\; \hs^{2} \;+\; \cdots \right) \right|^{2} \nonumber \\
 & = & \tau_{-}^{4}\,(1 - \tau_{+}^{2})^{2}\;|1 \;-\; \hs|^{-2}.
\label{eq:R_counter_plus}
\end{eqnarray}
Second, the recirculation coefficient (\ref{eq:recir_plus}) enters into the conversion coefficient ${\sf C}_{-}$ at lowest order because of the interference between the zeroth-order conversion amplitude $-\,\beta_{a}^{*}$ and the one-circuit amplitude $(\tau_{a}\,\beta_{d}\,\beta_{c}\,\beta_{b}\,e^{i\Phi})\,\tau_{a}$ (see the boxed expression below region $a$ in Fig.~\ref{fig:neg_B}). Each additional circuit around the quadrangle adds successive powers of $\hs$ to the one-circuit amplitude. Hence, the net conversion coefficient ${\sf C}_{-} \equiv |B_{- out}^{a}|^{2}/|A_{in}^{a}|^{2}$ to the downward $k$-propagating wave is 
\begin{eqnarray}
{\sf C}_{-} & = & \left| -\,\beta_{a}^{*} + \left( \tau_{a}\,\beta_{d}\,\beta_{c}\,\beta_{b}\,e^{i\Phi}\right)\,\tau_{a} \left( 1 + \hs + \hs^{2} + \cdots \right) \right|^{2} \nonumber \\
 & = & (1 - \tau_{-}^{2}) \;+\; |1 \;-\; \hs|^{-2} \left[\; \tau_{-}^{4}\,(1 - \tau_{+}^{2})^{2}\,(1 - \tau_{-}^{2}) 
\right. \nonumber \\
 &  &\left.-\; \tau_{-}^{2}\,[(\hs + \hs^{*}) - 2\,|\hs|^{2}] \;\right],
\label{eq:C2_counter_plus}
\end{eqnarray}
where we used $|-a^{*} + b|^{2} = |a|^{2} + |b|^{2} - (a^{*}b^{*} + ab)$. The modular-eikonal coefficients (\ref{eq:T_counter_plus})-(\ref{eq:C2_counter_plus}) depend on the model parameters $\sigma$ and $|\ov{\eta}_{\pm}|$ but not the propagation parameter $\gamma$, which represents a simplifying property of the symmetric-quadrangle model used here. They also satisfy the energy-flux conservation law 
\begin{equation}
1 \;=\; {\sf T} \;+\; {\sf C}_{+} \;+\; {\sf C}_{-} \;+\; {\sf R},
\label{eq:energy_counter}
\end{equation}
which follows from Eq.~(\ref{eq:action_con}).

\begin{figure}
\epsfysize=2.5in
\epsfbox{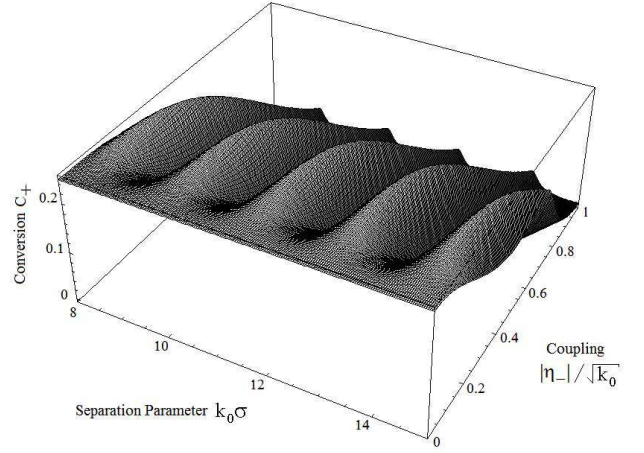}
\caption{Conversion coefficient (\ref{eq:C1_counter_plus}) as a function of $k_{0}\sigma$ and $|\eta_{-}|/\sqrt{k_{0}}$, for $\gamma = 1$ and 
$|\ov{\eta}_{+}| = [\ln(2)/2\pi]^{\frac{1}{2}}$. At fixed coupling constant $|\ov{\eta}_{-}|$, the conversion coefficient ${\sf C}_{+}$ exhibits interference effects as the separation parameter $k_{0}\sigma$ is varied and the interference phase (\ref{eq:Psi_counter}) passes through integer multiples of $\pi$.}
\label{fig:counter_pos_conv}
\end{figure}

Figure \ref{fig:counter_pos_conv} shows a plot of the conversion coefficient (\ref{eq:C1_counter_plus}) as a function of the separation parameter $k_{0}\sigma$ and the coupling constant $|\ov{\eta}_{-}|$, for $\gamma = 1$ and $|\ov{\eta}_{+}| = [\ln(2)/2\pi]^{\frac{1}{2}}$ (where the standard Budden conversion coefficient reaches its maximum at $\frac{1}{4}$); the other coefficients $({\sf T}, {\sf C}_{-}, {\sf R})$ exhibit similar interference effects while satisfying the conservation law (\ref{eq:energy_counter}). We make three remarks concerning our modular-eikonal coefficients (\ref{eq:T_counter_plus})-(\ref{eq:C2_counter_plus}). First, the standard Budden coefficients \cite{TK_93} are recovered from Eqs.~(\ref{eq:T_counter_plus})-(\ref{eq:C2_counter_plus}) with $\eta_{-} = 0$ (so that $\tau_{-} = 1$ and $\beta_{-} = 0 = \hs$). Second, as $|\eta_{-}| \rightarrow \infty$, we find ${\sf C}_{-} \rightarrow 1$ and $({\sf T}, {\sf C}_{+}, {\sf R}) \rightarrow 0$ as can be seen in Figure \ref{fig:counter_pos_conv} (where ${\sf C}_{+}$ becomes very small as $|\ov{\eta}_{-}| > 1$). Third, the use of an asymmetric-quadrangle model would only shift the minima and maxima in the modular-eikonal coefficients and preserve the interference effects. 

We note that all interference effects disappear when the order of the two counter-propagating waves in Fig.~\ref{fig:neg_B} is reversed (i.e., we assign negative values to the separation parameter $k_{0}\sigma$) since recirculation is no longer possible. In this case, the coefficients (\ref{eq:T_counter_plus})-(\ref{eq:C2_counter_plus}) simply become ${\sf T} = \tau_{+}^{2}\tau_{-}^{2}$, ${\sf C}_{+} = \tau_{+}^{2}\,(1 - \tau_{+}^{2})$, ${\sf R} = (1 - \tau_{+}^{2})^{2}$, and ${\sf C}_{-} = \tau_{+}^{2}\,(1 - \tau_{-}^{2})$, respectively, which still satisfy the conservation law (\ref{eq:energy_counter}). This reversed scenario may be viewed as a combination of a standard double-conversion Budden problem, with coefficients ${\sf T}_{B} = {\sf T} + {\sf C}_{-}$, ${\sf C}_{B} = {\sf C}_{+}$, and ${\sf R}_{B} = {\sf R}$, and an isolated single-conversion problem.

\begin{figure}
\epsfysize=2.5in
\epsfbox{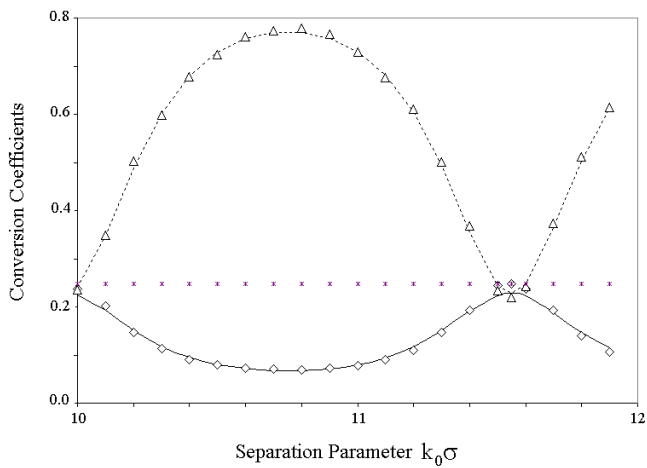}
\caption{Conversion coefficients ${\sf C}_{\pm}$ versus separation parameter $k_{0}\sigma$ for $\gamma = 1$ and $|\ov{\eta}_{+}| = |\ov{\eta}_{-}| = 
[\ln(2)/2\pi]^{\frac{1}{2}}$. The solid and dotted lines represent the theoretical formulas (\ref{eq:C1_counter_plus}) and (\ref{eq:C2_counter_plus}), respectively, and the symbols $\diamondsuit$ and $\triangle$  represent numerical results for ${\sf C}_{+}$ and ${\sf C}_{-}$, respectively. The standard Budden conversion ${\sf C}_{B}$ corresponding to $\eta_{-} = 0$ is also shown (represented by a horizontal set of $\times$ symbols).}
\label{fig:conversion}
\end{figure}

We now compare the analytical results presented above with the direct numerical integration of the coupled-wave differential equations (\ref{eq:ODE_Budden}) with $\gamma = 1$, $|\ov{\eta}_{+}| = [\ln(2)/2\pi]^{\frac{1}{2}} = |\ov{\eta}_{-}|$, and the boundary conditions corresponding to only a left-propagating $A$-wave with eikonal amplitude $A_{0}$ at $x = -\,\infty$ (and no other waves). With these boundary conditions, the energy-flux conservation law (\ref{eq:action_con}) yields 
\[ -\,k_{0}\,|A_{0}|^{2} \;=\; k_{0}\left(|A_{R}(+)|^{2} - |A_{L}(+)|^{2}\right) \;+\; \sum_{j = \pm}\;J_{j}(+), \]
where $A_{R,L}(+)$ denote the eikonal amplitudes of the right ($R$) and left ($L$) propagating $A$-waves at $x = +\,
\infty$. By comparing this equation with Eq.~(\ref{eq:energy_counter}), we define the numerical coefficients ${\sf T} = 
|A_{0}|^{2}/|A_{L}(+)|^{2}$, ${\sf C}_{\pm} = J_{\pm}(+)/k_{0}|A_{L}(+)|^{2}$, and ${\sf R} = |A_{R}(+)|^{2}/
|A_{L}(+)|^{2}$. The conversion coefficients ${\sf C}_{\pm}$ are shown in Fig.~\ref{fig:conversion} for $\gamma = 1$ and 
$|\ov{\eta}_{+}| = |\ov{\eta}_{-}| = \frac{1}{\sqrt{8}}$ (the latter value corresponds to the strongest interference effects observed in Fig.~\ref{fig:counter_pos_conv}), where excellent agreement is found between numerical results and the analytical conversion coefficients (\ref{eq:C1_counter_plus}) and (\ref{eq:C2_counter_plus}). The value of the conversion coefficient ${\sf C}_{B}$ corresponding to the standard Budden problem (with 
$\eta_{-} = 0$) is also shown and it is clearly seen that the presence of the third (counter-propagating) wave reduces the conversion to the bulk-ion wave, i.e., ${\sf C}_{+} < C_{B}$. We note that the modular-eikonal analysis \cite{Liang,Brizard_1} is valid only when the area (approximately $4\,k_{0}\sigma$) enclosed by the quadrangle in Fig.~\ref{fig:neg_B} is large enough compared to $2\pi$ (i.e., the separation $\sigma$ is large compared to the wavelength).

We conclude that the presence of an energetic-particle population may interfere with RF heating and current-drive processes in high-temperature magnetized plasmas. The simple one-dimensional model presented here for an extended Budden problem involving three coupled positive-energy waves demonstrates these interference effects and our numerical analysis validates the use of the modular-eikonal approach. The co-propagating and counter-propagating scenarios associated with a negative-energy wave \cite{Brizard_1,Brizard_2} supported by an inverted population of energetic-particles allow $\tau_{-}= \exp(+\,\pi |\ov{\eta}_{-}|^{2}) > 1$ and yield negative conversion coefficients ${\sf C}_{-}$, which can be expected to give bulk-ion conversion coefficients ${\sf C}_{+}$ that could exceed unity even as the conservation law (\ref{eq:energy_counter}) is preserved. We will investigate this potentially new alpha-channeling process, as well as develop more realistic models, in future work.

This work was supported by U.S.~DoE under grant No.~DE-AC03-76SFOO098. One of us (E.R.T.) also acknowledges support from U.S.~DoE under grant No.~DE-FG02-96ER54344 and NSF-DoE under contract No.~DE-FG02-06ER54885.

\end{document}